# Exclusive Generation of Single-Atom Sulfur for Ultrahigh Quality Monolayer MoS$_2$ Growth

Yunhao Zhang, [1,#] Jingwei Wang,[1,#] Yumo Chen,[1] Xian Wu,[1] Junyang Tan,[1] Jiarong Liu,[1] Huiyu Nong,[1] Liqiong He,[1] Qinke Wu,[1] Guangmin Zhou,[1] Xiaolong Zou,[1] Bilu Liu[1],*

[1] Shenzhen Geim Graphene Center, Shenzhen Key Laboratory of Advanced Layered Materials for Value-added Applications, Tsinghua-Berkeley Shenzhen Institute and Institute of Materials Research, Tsinghua Shenzhen International Graduate School, Tsinghua University, Shenzhen 518055, PR China

[#] These authors contributed equally.
Corresponding author: bilu.liu@sz.tsinghua.edu.cn

**Abstract**

Preparation of high-quality two-dimensional (2D) transition metal dichalcogenides (TMDCs) is the precondition for realizing their applications. However, the synthesized 2D TMDCs (*e.g.*, MoS$_2$) crystals suffer from low quality due to the massive defects formed during the growth. Here, we report the single-atom sulfur (S$_1$) as a highly reactive sulfur species to grow ultrahigh-quality monolayer MoS$_2$. Derived from battery waste, the sulfurized polyacrylonitrile (SPAN) is found to be exclusive and efficient in releasing S$_1$. The monolayer MoS$_2$ prepared by SPAN exhibits an ultralow defect density of ~7×10$^{12}$ cm$^{−2}$ and the narrowest photoluminescence (PL) emission peak with full-width at half-maximum of ~47.11 meV at room temperature. Moreover, the statistical resonance Raman and low-temperature PL results further verify the significantly lower defect density and higher optical quality of SPAN-grown MoS$_2$ than the conventional S-powder-grown samples. This work provides an effective approach for preparing ultrahigh-quality 2D single crystals, facilitating their industrial applications.

**Introduction**

Two-dimensional (2D) transition metal dichalcogenides (TMDCs) have garnered increasing attention owing to their atomically thin thickness and superior electronic performance.[1-5] For example, $MoS_2$ is recognized as a favorable candidate for the ultimate transistor scaling such as in memristors[6], photodetectors[7], piezotronics[8], and integrated circuits[9]. Synthesis of 2D TMDC crystals with high quality is a determinant for realizing their applications. Among various preparation methods, chemical vapor deposition (CVD) is the most promising due to high scalability, fine controllability, and compatibility with semiconductor manufacturing.[10-15] However, the TMDCs prepared by CVD usually suffer from abundant defects dominated by monosulfur vacancies, which are orders of magnitude higher than commercial Si and III-V compound semiconductors.[16, 17] These defects are important scattering sources for carriers and cause the degradation in electrical performance.[18, 19] It is reported that the electron mobility in monolayer $MoS_2$ decreased by two orders of magnitude as the sulfur defect density increased from $1\times10^{12}$ $cm^{-2}$ to $3\times10^{13}$ $cm^{-2}$.[20] Therefore, reducing the defect density in TMDCs is crucial.

From a chemical reaction point of view, during high-temperature vapor phase deposition, the defect formation in $MoS_2$ is a dynamic equilibrium process including the escape of sulfur atoms from the lattice and the vacancy repairing by gaseous sulfur species. The escaped sulfur atoms increase with temperature, and the defect repair depends on the activity of the sulfur species. As a mostly used chalcogen precursor, sulfur powder generates large sulfur clusters (*e.g.*, $S_8$, $S_6$, $S_4$, and $S_2$) with low reactivity, which makes it difficult to repair sulfur defects during growth. Recently, Feng et al. used liquid thiol to synthesize 2D $MoS_2$, where thiol molecules act as sulfur suppliers and defect repairers.[21] Jin et al. reported that the $Na_2SO_4$ effectively regulates the generation and mass flux of $H_2S$ by controlling the decomposition of $Na_2SO_4$ and obtained monolayer $MoS_2$ with low defect density.[22] In another work, Zuo et al. used ZnS to grow high-quality TMDCs and their alloys at relatively high temperatures.[23] Although many efforts have been made to improve the quality of TMDCs, the intrinsic interactions between sulfur species and defects remain ambiguous. This restricts the discovery of sulfur species with high chemical reactivity to suppress the defects. Therefore, developing efficient precursors for exclusively generating the corresponding sulfur species becomes an urgent issue.

Herein, we demonstrate that the single-atom sulfur ($S_1$) is a highly reactive sulfur species that can remarkably suppress the sulfur defects in $MoS_2$ based on density functional theory (DFT) calculations. We found that the sulfurized polyacrylonitrile (SPAN) from lithium-sulfur (Li-S) battery waste can exclusively produce $S_1$, which was used as the sulfur precursor to synthesize monolayer TMDCs. The $MoS_2$ grown by SPAN has an ultralow defect density of ~$7\times10^{12}$ $cm^{-2}$ and the smallest photoluminescence (PL) full-width at half-maximum (FWHM) of ~47.11 meV. Furthermore, the low-temperature PL and resonance Raman spectroscopy results both demonstrate that the $MoS_2$ samples grown by SPAN exhibit lower defect density and higher optical performance than the samples grown by conventional S powder.

**Results and discussions**

We performed DFT calculations to evaluate the interaction strength between different sulfur species and sulfur defects, through adsorption energies defined as eq. S1-2. As shown in Figure 1, the commonly utilized sulfur species including sulfur powder (*i.e.*, $S_2$ at CVD growth temperature), $H_2S$, and thiol (R-SH) possess relatively lower adsorption energies with sulfur defects in $MoS_2$, while $S_1$ exhibits the highest adsorption energy. This result suggests a strong thermodynamic preference for $S_1$ to combine with defects. Therefore, if $S_1$ is used as the chalcogen species, the sulfur defects can be effectively repaired and suppressed during growth, which is beneficial for the preparation of TMDCs with ultrahigh quality. Now the key challenge is how to exclusively and efficiently generate $S_1$ during growth.

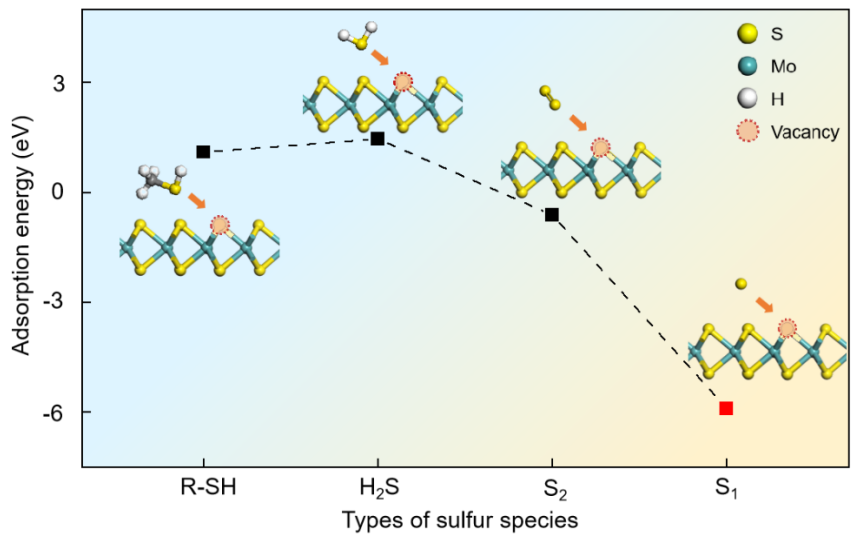

**Figure 1. DFT analysis on the adsorption energies of different sulfur species and the sulfur defect in MoS₂.** The insets show models of sulfur vacancies repaired by different sulfur species (*e.g.*, thiol, $H_2S$, $S_2$, and $S_1$).

Inspired by Li-S battery research, we came up with the molecule SPAN, in which sulfur (S) is covalently bonded to polyacrylonitrile (PAN) backbones. SPAN has been widely used as the cathode material to effectively reduce the shuttling of lithium polysulfides in Li-S batteries.[24-26] During discharge, the cross-linked S-S bond undergoes cleavage and lithiation on an aromatic structure, and separates the S atoms, which may present a promising sulfur precursor to generate $S_1$ for growing MoS₂. DFT calculations were conducted to simulate the release of $S_1$ and $S_2$ from both SPAN and S powder ($S_8$) precursors. As illustrated in Figure 2a, the release of $S_1$ from SPAN (I) undergoes a consecutive process. Initially, the breaking of the S-C bond occurs driven by the S-Li electrostatic force (II), followed by the formation of Li-C electrostatic interactions concurrent with the release of $S_1$ (III). Lithium atoms play a pivotal role in stabilizing the SPAN structure, thereby leading to a notably lower energy barrier for $S_1$ release (2.032 eV) compared to the S powder case (4.311 eV) (Figure 2b, c). Additionally, the energy barrier for $S_1$ release from SPAN is substantially lower than that of $S_2$ release (Figure S1), indicating a high selectivity toward $S_1$ during the thermal decomposition process. From the experimental point of view, the release of $S_1$ from SPAN can be verified by the TGA-mass measurement. Figure 2d and Figure S2 exhibit the proportion of each component from $S_1$ to $S_8$ when heating SPAN at 400 °C, which confirms the exclusive release of $S_1$ from SPAN. Note that the discharge process of SPAN is crucial for the release of $S_1$ (Figure S3). Overall, the above theory and experimental results both show that SPAN is a superior precursor that can selectively generate $S_1$.

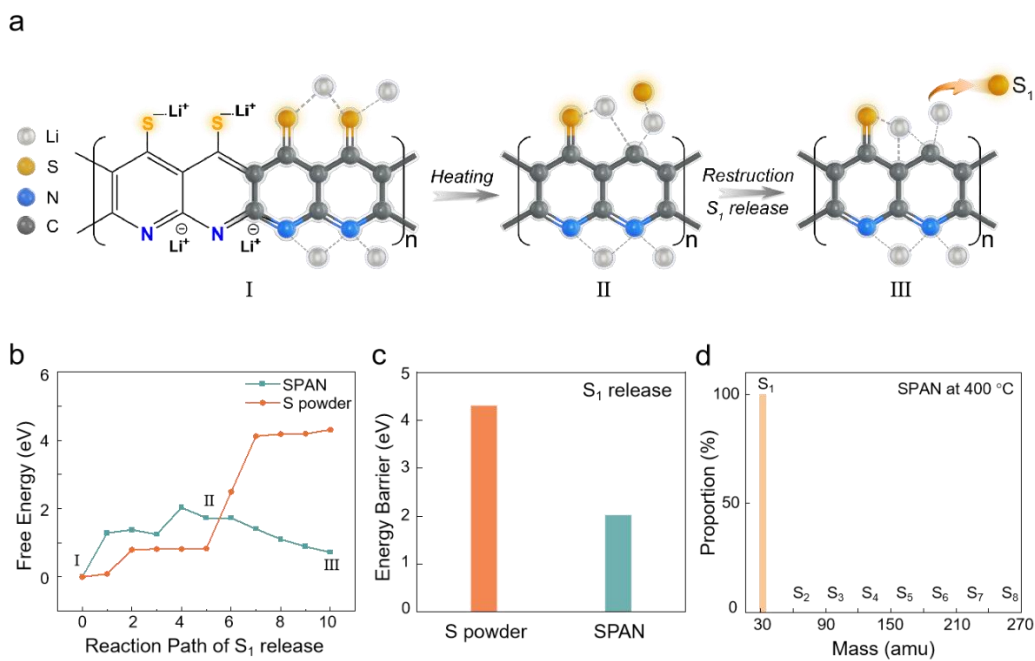

**Figure 2. Mechanism of $S_1$ generation.** (a) Schematic diagram of $S_1$ releasing process from SPAN. I represents the initial state of SPAN, followed by the SPAN restruction process (II). III is the final state and the release of $S_1$. (b) Calculated free energy for the $S_1$ release from S powder and SPAN, respectively. (c) Comparison of energy barriers for $S_1$ release from S powder and SPAN. (d) The proportion of each component from $S_1$ to $S_8$ when heating SPAN at 400 °C. The results are from the TGA-mass spectrum of SPAN in Figure S2.

Then we used the SPAN as the sulfur precursor to grow monolayer $MoS_2$ to test the above speculations, where the growth is illustrated in Figure 3a. The SPAN were obtained from the Li-S battery waste and their chemical composition were confirmed by X-ray photoelectron spectroscopy (XPS) in Figure S4,5. During growth, the heating temperature of SPAN precursor is set to 400°C since $S_1$ exhibits the highest release rate and selectivity at this temperature (Figure S2). The as-grown monolayer $MoS_2$ exhibit high uniformity with an average size of ~41 μm (Figure 3b-d and S6). The high-resolution atomic force microscopy (AFM) image in Figure 3e and the XPS results in Figure S7 indicate the formation of pure $MoS_2$ with nearly perfect hexagonal lattice. Furthermore, this preparation method is also applicable to the growth of high-quality and uniform $WS_2$ (Figure S8-10).

To evaluate the crystal quality of the samples, we carry out the aberration-corrected high-angle annular dark-field scanning transmission electron microscopy (HAADF-STEM) characterizations to measure the defect density of the monolayer $MoS_2$. Figure 3f and Figure S11 show the STEM images obtained from random regions across a large scale (~1000 $nm^2$), where only a very few sulfur defects are visualized without any presence of Mo vacancy. Owing to the Z-contrast nature of the corresponding HAADF-STEM image ((height profile in Figure 3g), the monosulfur defect could be distinguished from the brighter $S_2$ sites by the image contrast. The average density of sulfur defect was calculated to be ~7×$10^{12}$ $cm^{-2}$, which is the lowest value reported so far based on TEM measurements with a statistical area greater than 500 $nm^2$ (Figure 3h). [27-30] This value is also close to the theoretical optimization target.[18]

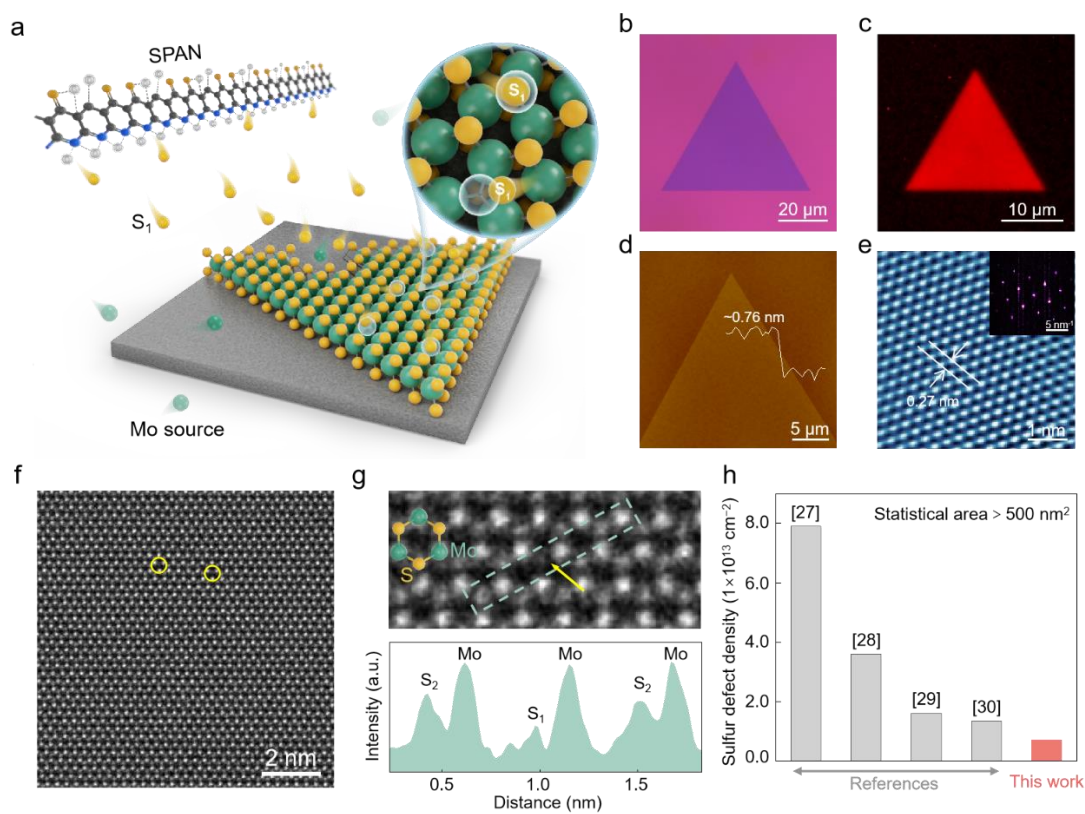

**Figure 3. CVD growth of monolayer MoS$_2$ by S$_1$ supply released from SPAN**. (a) Schematic of S$_1$ release from SPAN for the growth of MoS$_2$, where S$_1$ acts as the sulfur supplier and defect repairer. (b) Typical optical image, (c) fluorescence image, (d) AFM image, and (e) zoom-in high-resolution AFM image of the as-grown MoS$_2$. The inset of Figure 3e is the FFT image measured by high-resolution AFM. (f) HAADF-STEM image of SPAN-grown MoS$_2$. The sulfur defects are highlighted by the yellow circles. (g) Zoom-in image (top) of MoS$_2$ sample with monosulfur defect and its height profile (bottom) along the dashed green rectangle. (h) Comparison of sulfur defect density of SPAN-grown MoS$_2$ with other TEM statistic results with a measured area greater than 500 nm$^2$. [27-30]

To further investigate the optical quality of the MoS$_2$, we performed PL measurements and resonance Raman spectroscopy to compare the MoS$_2$ samples grown by SPAN and S powder. The room-temperature PL spectrum (Figure 4a) suggests that the MoS$_2$ grown by SPAN exhibits higher PL intensity and narrower FWHM, confirming their superior optical quality compared with the samples grown by S powder. More generally, 20 PL spectra from different monolayer flakes are obtained for each type of samples (Figure 4b), which shows all the SPAN-grown MoS$_2$ samples have high and uniform optical performance. It is notable that the SPAN-grown MoS$_2$ also reveals a sharp PL emission with an FWHM of ~47.11 meV, which is the narrowest value among the previous works (Figure 4a and Figure 4c).[10, 13, 23, 31-36] Furthermore, the low-temperature PL spectra at 80 K of MoS$_2$ samples grown via S powder and SPAN are compared in Figure 4d and Figure S12, which can be fitted into four peaks. The two high energy modes ascribe to neutral A exciton ($X_A$) and B exciton ($X_B$) accompanied by a negatively charged trion peak ($X_T$), while the lower energy peak is attributed to the defects. The MoS$_2$ grown by SPAN shows a weaker defect peak than that of the S powder case, revealing their ultrahigh quality consistent with the STEM results. Figure 4e shows the typical resonance Raman spectra of SPAN-grown and S-powder-grown MoS$_2$ excited by 633 nm wavelengths, where the characteristic LA(M) peak at ~227 cm$^{-1}$ is the defect-induced mode.[37-39] As shown in Figure 4e, the significantly smaller intensity of the LA(M) peak of the SPAN-grown MoS$_2$ reveals their lower defect density than the S-powder-grown samples. Furthermore, we have obtained the statistics of 50 Raman spectra for different monolayer flakes of each type of sample (Figure 4f), indicating the consistently high quality of SPAN-grown MoS$_2$ samples. Moreover, the MoS$_2$ grown by undischarged SPAN exhibits lower quality than the discharged SPAN-grown samples (Figure S13). Therefore, the above characterizations suggest the monolayer MoS$_2$ grown by SPAN possesses ultralow defect density and ultrahigh optical quality.

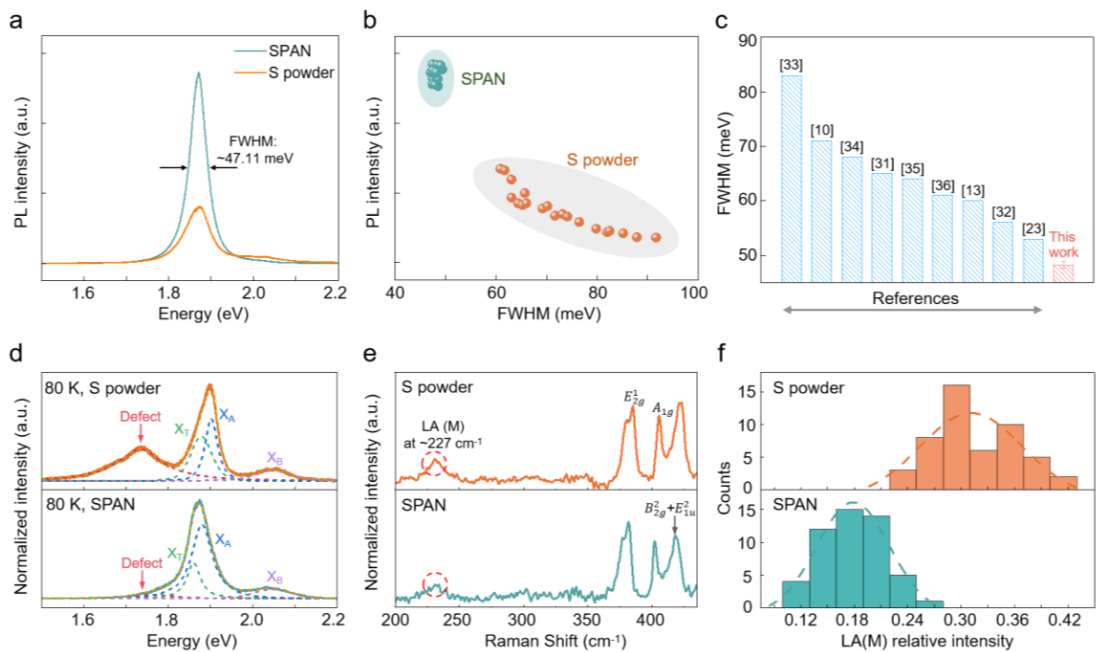

**Figure 4. Optical characterizations of the monolayer MoS$_2$ grown by SPAN and S powder.** (a) Room-temperature PL spectra of MoS$_2$ grown by SPAN and S powder. (b) Statistics of PL intensity and FWHM for SPAN-grown and S-powder-grown samples. (c) The FWHM of MoS$_2$ grown by SPAN compared with the previously reported work. [10, 13, 23, 31-36] (d) Low-temperature PL characterization at 80 K of MoS$_2$ samples grown via S powder and SPAN. (e) Resonance Raman results of MoS$_2$ grown by SPAN and S powder. The characteristic LA(M) peak at ~227 cm$^{-1}$ assigned to the longitudinal acoustic (LA) at the M point in the Brillouin zone, which is a defect-sensitive mode. (f) Statistic distribution of LA(M) peak intensity for SPAN-grown and S-powder-grown MoS$_2$. These data are obtained from different flakes.

## Conclusion

In summary, we verified that S$_1$ possesses the highest adsorption energy with sulfur vacancy among so-far reported sulfur species. Derived from Li-S battery waste, SPAN is found to be an ideal precursor to exclusively release S$_1$ without any other forms of sulfur species, as confirmed by DFT calculations and in situ mass spectroscopy results. Using SPAN as the sulfur precursor, we successfully synthesized the ultrahigh-quality monolayer

TMDCs. Taking MoS$_2$ as a representative, the microscopic and spectroscopic characterizations demonstrate that the as-grown monolayer MoS$_2$ samples possess an ultralow density of sulfur defect (~7×10$^{12}$ cm$^{-2}$) and ultrahigh optical quality with a PL FWHM of ~47.11 meV. Compared to the S-powder-grown samples, the MoS$_2$ grown by SPAN exhibits significantly lower defect density and higher optical performance. Our strategy not only enriches the understanding of the relationship between sulfur species and defects in MoS$_2$ but also provides a potential strategy to grow other high-quality 2D TMDCs.

**Supporting Information**

DFT calculations on the mechanism of S$_1$ and S$_2$ released from SPAN, thermogravimetric analysis (TGA) analysis and TGA-mass of SPAN precursor, DFT calculations on the mechanism of S$_1$ and S$_2$ released from raw SPAN, Photos of SPAN powder, XPS spectra results of SPAN, optical images of MoS$_2$ flakes and statistical analysis of their average size, XPS results of the monolayer MoS$_2$ grown by SPAN, optical images, AFM images and PL spectra of as-grown monolayer WS$_2$, optical images of WS$_2$ flakes and statistical analysis of their average size, Raman spectra of WS$_2$ grown by SPAN and S powder, HAADF-STEM images of SPAN-grown MoS$_2$ captured at different regions, low-temperature PL spectra of SPAN-grown MoS$_2$ at 80K.

**Notes**

The authors declare no competing financial interest.

**Acknowledgements**

The authors acknowledge the support by the National Key R&D Program (2022YFA1204301), the National Science Foundation of China for Distinguished Young Scholars (52125309), the National Natural Science Foundation of China (51991343,

51991340, 51920105002, and 52188101), the Natural Science Foundation of Guangdong Province of China (2023A1515011752), the Guangdong Basic and Applied Basic Research Foundation (2023A1515110411), the Shenzhen Basic Research Project (JCYJ20220818101014029, JCYJ20230807111619039), the Shenzhen Science and Technology Program (ZDSYS20230626091100001), the Shuimu Tsinghua Scholar Program of Tsinghua University (2022SM092), and the Tsinghua Shenzhen International Graduate School-Shenzhen Pengrui Young Faculty Program of Shenzhen Pengrui Foundation (No. SZPR2023002). The authors appreciate Prof. Yu Lei's discussions.

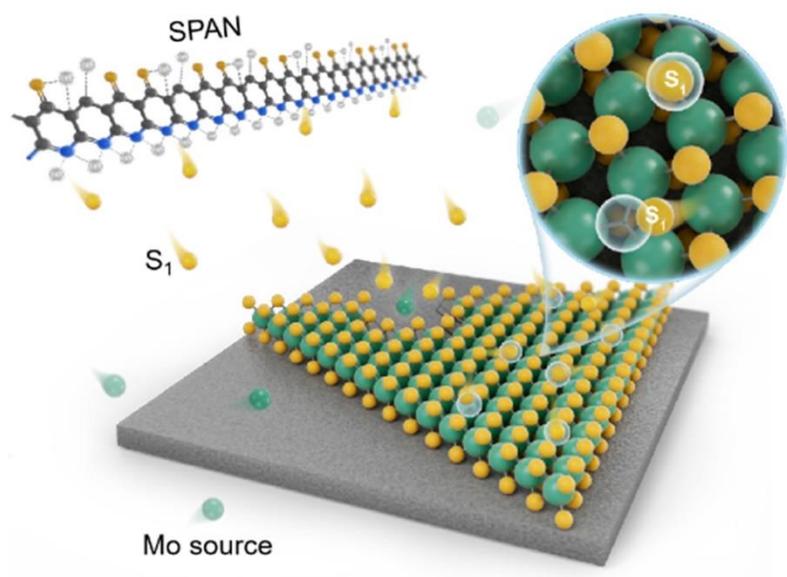

**TOC Graphic**